%% file: main.tex
\documentclass[10pt, conference, letterpaper]{IEEEtran}

\usepackage[ruled,noline]{algorithm2e}
\usepackage{amsfonts,amsmath,tikz}
\usepackage{comment}

\newtheorem{theorem}{Theorem}[section]

\newtheorem{lemma}[theorem]{Lemma}
\newtheorem{remark}{Remark}[section]
\newtheorem{definition}{Definition}[section]
\newtheorem{exmp}{Example}[section]

\newcommand{\greedy}{{\normalsize G}{\scriptsize REEDY}}
\newcommand{\onlinegreedy}{\normalsize{O}\scriptsize{NLINE}\normalsize{G}\scriptsize{REEDY}\normalsize{ }}
\newcommand{\randomonlinegreedy}{\normalsize{R}\scriptsize{ANDOM}\normalsize{O}\scriptsize{NLINE}\normalsize{G}\scriptsize{REEDY}\normalsize{ }}
\newcommand{\parallelalgo}{\normalsize{P}\scriptsize{ARALLEL}\normalsize{L}\scriptsize{OAD}\normalsize{B}\scriptsize{ALANCE}\normalsize{ }}

\newcommand{\mA}{\mathcal{A}}

\title{\LARGE\bf
Online Budgeted Repeated Matching
}
\author{\parbox{3 in}{ \centering Ajil Jalal\\
        Department of Electrical Engineering \\
        Indian Institute of Technology Madras\\
      	Chennai, India\\
        {\tt\small ee12b004@ee.iitm.ac.in}} 
		\hspace*{ 0.5 in}		
		\parbox{4 in}{\centering Rahul Vaze, Umang Bhaskar\\
        School of Technology and Computer Science\\
        Tata Institute of Fundamental Research\\
        Mumbai, India\\
        {\tt\small vaze@tcs.tifr.res.in umang@tifr.res.in}}
}

\begin{document}
\maketitle

\begin{abstract}A basic combinatorial online resource allocation problem is considered, where multiple servers have individual capacity constraints, and at each time slot, a set of jobs arrives, that have potentially different weights to different servers. At each time slot, a one-to-one matching has to be found between jobs and servers, subject to individual capacity constraints, in an online manner. The objective is to maximize the aggregate weight of jobs allotted to servers, summed across time slots and servers, subject to individual capacity constraints. This problem generalizes the well known adwords problem, and is also relevant for various other modern applications.
A simple greedy algorithm is shown to be $3$-competitive, whenever the weight of any edge is at most half of the corresponding server capacity. Moreover, a randomized version of the greedy algorithm is shown to be $6$-competitive for the unrestricted edge weights case. For parallel servers with small-weight jobs, we show that a load-balancing algorithm is near-optimal.	
\end{abstract}

\thispagestyle{empty}
\pagestyle{empty}
\input{introduction.tex}

%

\section{Problem Definition}
We are given a set $I$ of $n$ servers, where server $i$ has capacity $C_i$. We consider an \textit{online} scenario, in which at each time step $t \in \{1, \dots, T\}$, a set of jobs $J(t)$ and a set of edges $E(t)$ from servers $I$ to jobs $J(t)$ is revealed. Edges are weighted, and $w(e)$ for $e = (i,j)$ is the quantity of resources of server $i$ consumes if job $j$ is assigned to server $i$, or the weight of job $j$ on server $i$. In general, a job may have different weights on different servers, thus for distinct servers $i$ and $i'$, $w(i,j) \neq w(i',j)$. The entire set of jobs is $J = \cup_{t \leq T} J(t)$, and $E = \cup_{t \leq T} E(t)$. We also define $w(i,j) := w(e)$ where $e = (i,j)$, and for a set of edges $F$, define $W(F) := \sum_{e \in F} w(e)$, and $F(t) := F \cap E(t)$ as the set of edges incident to jobs in time step $t$. Define $G(t)$ as the bipartite graph $(I \cup J(t), E(t))$. A set of edges $F$ is \emph{feasible} if (i) $F(t)$ is a matching for all $t \leq T$, i.e., each server and job is connected to at most one job and one server, respectively, at each $t$, and (ii) the total weight of edges incident to each server is at most its capacity. We will also call a feasible set of edges an \emph{allocation}.

The Online Budgeted Repeated Matching (OBRM) problem is to pick matchings $M(t) \subseteq E(t)$ \emph{irrevocably} at each time step $t$ to maximize $W(\cup_{t \leq T} M(t))$, so that the weight of edges in $\cup_{t \leq T} M(t)$ incident to server $i$ is at most $C_i$. 

An optimal allocation for an instance of OBRM has maximum weight among all allocations. The \emph{competitive ratio} for an algorithm for the OBRM problem is defined as the minimum over all instances of the ratio of the weight of the allocation obtained by the algorithm, to the weight of the optimal allocation for the instance. For a \emph{randomized} algorithm, the competitive ratio is obtained by taking the numerator of the previous ratio as the \emph{expected} weight of the allocation obtained by the algorithm. We use $\mu(\mA)$ to denote the competitive ratio for an algorithm $\mA$.

\section{Algorithms}
\begin{figure}[h!]
	\begin{center}
    \includegraphics[width=0.5\textwidth]{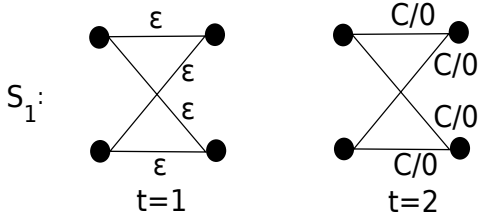}
    \caption{Illustration for example \ref{example: deterministic adversarial setting}.}
    \label{figure: illustrative example for bad performance}	
\end{center}	    
\end{figure}
\begin{exmp}
\label{example: deterministic adversarial setting}
We begin by illustrating via an example the difficulty in solving the OBRM problem. Consider Fig. \ref{figure: illustrative example for bad performance},  where the job weights are marked next to the corresponding edges in the graphs, and both servers have capacity $C$(assume $\epsilon\ll C$). 
The weights at time $t=2$ are chosen by the adversary as follows: if server $1/2$ has not been matched at $t=1$, then all the jobs that they can serve at $t=2$ have weight $0$, else the jobs to server $1/2$ have weight $C$. Thus, $S_1$ can be thought of as 4 different sequences based on the matching produced at $t=1$. 

Now, if any online algorithm assigns  a job of weight $\epsilon$ to any server at time $1$, then it would prevent it from accepting a job of weight $C$ at time $2$, whereas rejecting the job of weight $\epsilon$ would make sure that the algorithm cannot allocate any jobs if the weights in time $2$ are all zero, i.e., accepting/rejecting a job at time $1$ can cause an arbitrarily bad competitive ratio in the next time step. This makes the worst case competitive ratio as small as $\epsilon/C$ for all deterministic algorithms.


However, if we restrict the maximum weight of a job to be $\frac{C}{2}$, then every server can accept at least two jobs, and a deterministic algorithm can give a non-trivial competitive ratio even on adversarial sequences. Under this restriction, we propose an  \onlinegreedy algorithm that is shown to be $3$-competitive next.
\end{exmp}
\vspace*{0.1in}

In the discussion of the following algorithms, we use $M(t)$ to denote the set of edges selected by the algorithm in time step $t$, $A(t) := \cup_{\tau \le t} M(\tau)$, and $M_i(t)$ and $A_i(t)$ to denote the set of edges in $M(t)$ and $A(t)$ incident to server $i$.

\subsection{Deterministic Algorithm for Restricted Edge Weights}
\begin{definition}
Active server: The server $i$ is active at time step $t+1$ if the sum of the weights of edges assigned to it so far is at most half its capacity, i.e., $W(A_i(t))\leq\frac{1}{2}C_i$. 
\vspace*{0.05in}
\end{definition}

We will use $S$ to denote the set of active servers. We first describe an algorithm that will be used as an intermediary for our final online algorithms, \onlinegreedy and \randomonlinegreedy.

\subsubsection{\greedy}
The deterministic algorithm \greedy$\;$ takes as inputs a weighted bipartite graph $G$, as well as a set $S$ of active servers. \greedy$\;$  greedily picks edges from the bipartite graph $G$ to form a matching $M$. The algorithm only adds an edge to the matching if the server connected to it is active. 

\subsubsection{\onlinegreedy}
\begin{algorithm}[h]
\label{algorithm: Greedy}
\caption{\greedy($G,S$)}
\SetKwInOut{Input}{Input}\SetKwInOut{Output}{Output}
\Input{Weighted bipartite graph $G$, set of active servers $S$}
\Output{Matching $M$}
\Begin{
$M\leftarrow\emptyset$\\
\For{$e=(i,j)\in G$ in descending order of weight}{
	\If{($M\cup e$ is a matching) \textbf{AND} $\left(i\in S\right)$}{
		$M\leftarrow M\cup e$\\	
	}
	\vspace*{0.1in}
}
}
\Return{M}

\end{algorithm}

\begin{algorithm}[h]
\label{algorithm: Restricted Online Greedy}
\caption{\onlinegreedy}
\SetKwInOut{Input}{Input}\SetKwInOut{Output}{Output}
\Input{Server capacities $C_1,C_2,...,C_n$\\ Weighted bipartite graphs $G(t)$ for $t \leq T$, \\ such that $w(i,j) \le \frac{1}{2}C_{i} \, \forall i,j$}
\Output{Feasible allocation $A(T) = \cup_{t \leq T} M(t)$}
\Begin{
$S\leftarrow I $\\
$A_{i}(0)\leftarrow\emptyset\;\forall\; i\in I$\\
\For{$t\leftarrow1$ \KwTo $T$}{
	$M(t)\leftarrow$\greedy$(G(t),S)$\\
	$A(t) \leftarrow A(t-1) \cup M(t)$\\
	\For{$(i,j) \in M(t)$}{
		\If{$W\left(A_{i}(t)\right) > \frac{C_{i}}{2}$}{	
			$S\leftarrow S\setminus \{i\}$\\
		}	
	}	
}
}
\end{algorithm}

We present a deterministic algorithm \onlinegreedy that is $3$-competitive for the restricted weights case, where the weight of each edge incident to a server is at most half the server capacity, i.e., $w(i,j) \le \frac{1}{2}C_i$ for each server $i$ and job $j$. 

\onlinegreedy maintains a set of active servers $S$, along with sets $A_i(t)$ for each server $i$, where $A_i(t)$ is the set of edges selected that are incident to server $i$ until time $t$. At each time step $t$, \onlinegreedy calls \greedy$\;$and passes to it as input the weighted bipartite graph $G(t)$ along with the current set of active servers $S$. For each edge $(i,j) \in M(t)$, where $M(t)$ is the matching returned by \greedy, edge $(i,j)$ is added to the allocation $A_i(t)$. \onlinegreedy then checks if $W(A_i(t))> \frac{1}{2}C_i$, in which case server $i$ is no longer active and is removed from the set of active servers $S$ for next time slot. If a server $i$ is active at time $t$, i.e., $W(A_i(t-1))\leq\frac{1}{2}C_i$, and an edge $e$ is added to $A_i(t-1)$, then $W(A_i(t-1))$ increases by at most $\frac{1}{2}C_i$, and hence $W(A_i(t))\leq C_i$. Hence, assigning a job to an active server always results in a feasible allocation. Also, since \greedy$\;$ performs a matching at each time step, the degree constraints (one job/server is assigned to at most one server/job, respectively) are always satisfied. The algorithm continues either until $S=\emptyset$ or $t=T$.

\begin{remark}
We note that the restriction on edge weights is only used in proving the feasibility of the allocation obtained, and not in the proof of 3-competitiveness below. In particular, if the edge weights are unrestricted, the allocation obtained may violate the capacity constraints, but will be 3-competitive. 
\end{remark}

\vspace*{0.1in}
\begin{theorem}\label{theorem: online greedy theorem}
\onlinegreedy is $3$-competitive.
\end{theorem}
\begin{IEEEproof}
For each time step $t$, let $M(t)$ denote the matching produced by \onlinegreedy, and let $M^*(t)$ denote the corresponding matching given by the optimal offline algorithm. Let $A^*(t) = \cup_{\tau \leq t} M^*(\tau)$, and $A_i^*(t)$ is the set of edges to server $i$ in the optimal allocation until time $t$. Also, $A_i^*=A_i^*(T)$, $A_i=A_i(T)$, and $A = \cup_{i \in I} A_i$, $A^* = \cup_{i \in I} A_i^*$.

We say that an edge $e=(i,j)\in M^{*}(t)\setminus M(t)$, has been \textit{blocked} by a heavier weight edge $f\in M(t)$ if $w(f)\geq w(e)$ and $f$ shares a server vertex ($i$) or job vertex ($j$) with $e$. As $f$ has more weight than $e$, \greedy$\;$ would select it first in $M(t)$, and hence $e$ cannot be selected without violating matching constraints. For each edge $(i,j) \in M^{*}(t)\setminus M(t)$, there are three possible reasons why the edge was not selected by \onlinegreedy:
\begin{enumerate}
\item An edge $f=(i,j') \in M(t),\,j'\neq j$ \textit{blocks} $(i,j)$, i.e. server $i$ was matched to some job $j'$ by \greedy, such that $w(i,j')\geq w(i,j)$.


\item An edge $f=(i',j) \in M(t),\,i'\neq i$ \textit{blocks} $(i,j)$, i.e. job $j$ was matched to some server $i'$ by \greedy, such that $w(i',j)\geq w(i,j)$.


\item The server $i$ was inactive at time step $t$, i.e., $i\notin S$. 

\end{enumerate}

Let $E_{1}(t)$, $E_{2}(t)$ and $E_{3}(t)$ denote the set of edges in $M^*(t) \setminus M(t)$ that satisfy the first, second and third condition respectively. Clearly, $E_{1}(t)\cup E_{2}(t)\cup E_{3}(t) =M^{*}_{t}\setminus M_{t}$. \textit{Note:} No edge can satisfy the first and third condition simultaneously, as a server which is inactive at time $t$ cannot be matched to any job at time $t$. Therefore, $E_{1}(t)\cap E_{3}(t)=\emptyset$. However, in general, $E_{1}(t)\cap E_{2}(t)\neq\emptyset$ and $E_{2}(t)\cap E_{3}(t)\neq\emptyset$, as edges can satisfy conditions 1 and 2 or 2 and 3.

\vspace*{0.1in}

Let $S$ be the set of active servers at time $T+1$. For all servers $i, i\notin S$, since $W(A^*_i)\leq C_i$ and $W(A_i)>\frac{1}{2}C_i$, the allocation $A_i$ is a $\frac{1}{2}$ approximation to $A_i^*$, i.e.,
\begin{eqnarray}\label{equation: 3comp sum i not in S}
\sum_{i:i\notin S}\sum_{e\in A_i^*}w(e)< 2\sum_{i:i\notin S}\sum_{e\in A_i}w(e)\,.
\end{eqnarray}

\vspace*{0.1in}

Let $E_1=\cup_{t=1}^{T}E_1(t),E_2=\cup_{t=1}^{T}E_2(t),E_3=\cup_{i=1}^{T}E_3(t)$.  Define $E_1^S=\{e=(i,j)\in E_1\mid i\in S\},E_2^S=\{e=(i,j)\in E_2\mid i\in S\}$. Clearly, $E_1^S\cup E_2^S=\cup_{i:i\in S}\left(A_i^*\setminus A_i\right)$, as no edge $e=(i,j),i\in S$ can satisfy the third condition.

The edges $e\in E_1^S\cup E_2^S$ were not selected in the greedy allocation as they were blocked by edges of heavier weight from $A\setminus A^*$. The edges in the set $A\setminus A^*$ are of two types:
\begin{enumerate}
\item $f=(i,j)\in A_i\setminus A_i^*, i\in S$. As all edges $e=(i',j')\in E_1^S\cup E_2^S$ are such that $i'\in S$, $e$ was blocked either because $e$ and $f$ share a server vertex ($i=i'$) or they share a job vertex ($j=j'$). Thus, for every edge $f=(i,j)\in A_i\setminus A_i^*,i\in S$, there may exist at most two edges $e_1=(i,j'),e_2=(i',j)$ that are blocked by $f$, so that $e_1,e_2\in E_1^S \cup E_2^S$ and $w(f)\geq w(e_1),w(f)\geq w(e_2)$.

\item $g=(i,j)\in A_i\setminus A_i^*, i\notin S$. As all edges $e=(i',j')\in E_1^S\cup E_2^S$ are such that $i'\in S$, $e$ was blocked only because $g$ and $e$ share the same job vertex ($j=j'$) and $g$ was greedily picked first. Thus, for every edge $g=(i,j)\in A\setminus A^*,i\notin S$, there may exist at most one edge $e_1=(i',j)\in E_1^S\cup E_2^S$ that is blocked by $g$ and is such that $w(g)\geq w(e_1)$.
\end{enumerate}

As $f=(i,j)\in A_i\setminus A_i^*, i\in S$ can block at most two edges in $E_1^S \cup E_2^S$ and $g=(i,j)\in A_i\setminus A_i^*, i\notin S$ can block at most one edge in $E_1^S \cup E_2^S$,
\begin{displaymath}
\sum_{i:i\in S}\sum_{e\in A^*_i\setminus A_i}w(e)=\sum_{e\in E_1^S \cup E_2^S}w(e)
\end{displaymath}
\begin{equation}\label{equation: 3comp sum i in S}
\leq2\sum_{i:i\in S}\sum_{f\in A_i\setminus A_i^*}w(f)+\sum_{i:i\notin S}\sum_{g\in A_i\setminus A^*_i}w(g)\,.
\end{equation}



Adding (\ref{equation: 3comp sum i not in S}), (\ref{equation: 3comp sum i in S}),
\begin{eqnarray}
\sum_{i:i\notin S}\sum_{e\in A_i^*}w(e) + \sum_{i:i\in S}\sum_{e\in A^*_i\setminus A_i}w(e)\leq 2\sum_{i:i\notin S}\sum_{e\in A_i}w(e) +\nonumber\\
2\sum_{i:i\in S}\sum_{f\in A_i\setminus A_i^*}w(f) + \sum_{i:i\notin S}\sum_{g\in A_i\setminus A^*_i}w(g).\nonumber
\end{eqnarray}

Adding $\sum_{i:i\in S}\sum_{e\in A_i\cap A_i^*}w(e)$ to LHS and RHS,
\begin{eqnarray}
\sum_{i:i\notin S}\sum_{e\in A_i^*}w(e) +\sum_{i:i\in S}\sum_{e\in A^*_i}w(e)\leq\sum_{i:i\in S}\sum_{e\in A_i\cap A_i^*}w(e)&&\nonumber\\
+2\sum_{i:i\in S}\sum_{f\in A_i\setminus A_i^*}w(f)+3\sum_{i:i\notin S}\sum_{g\in A_i}w(g).&&\nonumber
\end{eqnarray}

Simplifying,
\begin{equation}\label{equation: 3comp final equation}
\sum_{i\in I}\sum_{e\in A_i^*}w(e)\leq 3\sum_{i\in I}\sum_{e\in A_i}w(e).\\
\end{equation}
\vspace*{0.1in}
\end{IEEEproof}

\begin{figure}[h!]
	\begin{center}
    \includegraphics[width=0.5\textwidth]{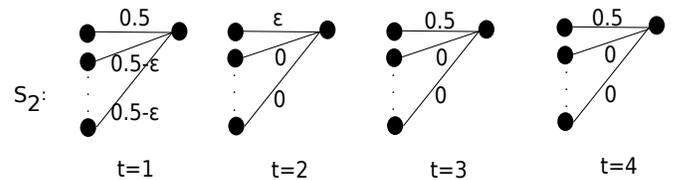}
    \caption{Illustration for example \ref{example: example for tightness of analysis}}
    \label{figure: illustrative example for tightness of onlinegreedy}	
\end{center}	    
\end{figure}

\begin{exmp}\label{example: example for tightness of analysis}
This example is used to show the tightness of analysis for \textbf{Theorem \ref{theorem: online greedy theorem}}. There are $n$ servers with capacity 1. The sequence of jobs is illustrated in Fig. \ref{figure: illustrative example for tightness of onlinegreedy}. At $t=1$, only the edge to server 1 has weight $0.5$, all other edges have weight $(0.5-\epsilon)$. At $t=2$, only the edge to server 1 has weight $\epsilon$, all other edges have weight $0$. At $t=3,4$ only the edge to server 1 has weight $0.5$, all other edges have weight 0. \onlinegreedy assigns the job at $t=1,2$ to server 1, and can't assign any more jobs at $t=3,4$, as server 1 is not active during those time slots, and the total weight of the allocation by \onlinegreedy is $0.5+\epsilon$. The optimal allocation would be to assign the job $(0.5-\epsilon)$ at $t=1$ to server 2, and then assign the jobs at time slot $t=3,4$ to server 1, so that the optimal weight allocation is $(1.5-\epsilon)$. Hence \onlinegreedy is a $\frac{1}{3}$-approximation, and this infinite family of instances shows that the analysis of the algorithm is tight.
\end{exmp}
\vspace*{0.1in}
\begin{remark}\label{remark: general competitive ratio of onlinegreedy}
In the more general case, where edge weights are restricted to be at most $\alpha \ (\leq 1)$ times the corresponding server capacities, i.e., if $w(i,j)\leq\alpha C_i\,\forall\,i,j$, the following modification of \onlinegreedy makes it $\left(1+\frac{1}{1-\alpha}\right)$-competitive. Instead of removing a server $i$ from the set of active servers $S$ when $W(A_i(t))>\frac{1}{2}C_i$, if we remove it when $W(A_i(t))>(1-\alpha)C_i$, then (\ref{equation: 3comp sum i not in S}) can be changed to
\begin{eqnarray}
\sum_{i:i\notin S}\sum_{e\in A_i^*}w(e)&<&\left(\dfrac{1}{1-\alpha}\right)\sum_{i:i\notin S}\sum_{e\in A_i}w(e).\nonumber
\end{eqnarray}
The rest of the proof follows directly to give a $\left(1+\frac{1}{1-\alpha}\right)$-competitive algorithm. Clearly, as $\alpha\rightarrow 1$, the competitive ratio tends to 0, and \onlinegreedy will fail, as expected from Example \ref{example: deterministic adversarial setting}. To handle the case of unrestricted job weights, in the next subsection, we present a randomized algorithm \randomonlinegreedy which is $6-$competitive.
\end{remark}

\subsection{Randomized Algorithm for Unrestricted Edge Weights}
\begin{algorithm}[h]
\label{algorithm: Unrestricted Random Online Greedy}
\caption{\randomonlinegreedy}
\SetKwInOut{Input}{Input}\SetKwInOut{Output}{Output}
\Input{Server capacities $C_1,C_2,...,C_n$\\ Weighted bipartite graph $G(t)$ for $t\leq T$,  such that $w(i,j)\leq C_i\,\forall\,i,j$}
\Output{Random feasible allocation $A=\cup_{i\in I}A_i$}
\Begin{
$S\leftarrow I$\\
$S_{1},\,S_{2}\leftarrow\emptyset$\\
$A_{i}(0),\;B_{i}(0)\leftarrow\emptyset\;\forall\; i\in I$\\
\vspace*{0.05in}
\text{// For each server}

\For{$k\leftarrow 1$ \KwTo $n$}{
	\vspace*{0.05in}
	$v_{k}\sim Bernoulli(\frac{1}{2})$\\
	\vspace*{0.05in}
	\eIf{$v_{k}=1$}{
		$S_{1}\leftarrow S_{1}\cup \{k\}$ \ \ \  \text{// accept only heavy jobs}\\
	}
	{$S_{2}\leftarrow S_{2}\cup \{k\}$ \ \ \  \text{// accept only light jobs} \\} 
}
\vspace*{0.05in}
\For{$t\leftarrow1$ \KwTo $T$}{

	\vspace*{0.05in}
	$M(t)\leftarrow$\greedy$(G(t),S)$\\
	
	\vspace*{0.05in}
	\For{$e=(i,j)\in M(t)$}{
		\vspace*{0.05in}		
		$B_{i}(t)\leftarrow B_{i}(t-1)\cup \{e\}$\\
		\vspace*{0.05in}		
		\If{$W\left(B_{i}(t)\right) > \frac{C_{i}}{2}$}{
			\vspace*{0.05in}	
			$S\leftarrow S\setminus \{i\}$\\	
		}
		\vspace*{0.05in}
		\If{$\left(i\in S_{1}\;\textbf{AND}\;w(i,j)> \frac{C_{i}}{2}\right)$ \textbf{OR} $\left(i\in S_{2}\;\textbf{AND}\;w(i,j)\leq \frac{C_{i}}{2}\right)$}{
			\vspace*{0.05in}			
			$A_{i}(t)\leftarrow A_{i}(t-1)\cup \{e\}$\\
		}		
	}
}
}
\end{algorithm}

Next, we present a randomized online algorithm \randomonlinegreedy that is $6-$competitive for the general case of unrestricted edge weights against an oblivious adversary that determines the input before the random coin flips. 
Note that while $w(i,j)$ can be unbounded, any edge such that $w(i,j)>C_i$ will be ignored as it can never be allocated to server $i$.
\vspace*{0.1in}
\begin{definition}
An edge $e=(i,j)$ that satisfies $\frac{C_i}{2} < w(i,j)\leq C_{i}$ is called a \textit{heavy edge} and the corresponding job is called a \textit{heavy job} for that server. In other words, the weight of a heavy edge $(i,j)$ connected to a server $i$ is at least half the server's initial capacity. An edge that is not heavy is called \textit{light}, and the corresponding job is called {\it light} for that server.
\end{definition}
\vspace*{0.1in}

At the start of the algorithm, an unbiased coin is flipped for each server $i$. If heads, then server $i$ is added to set $S_1$, else it is added to set $S_2$. If server $i \in S_1$, it can only accept jobs corresponding to heavy edges, while if $i\in S_2$, it can only accept jobs corresponding to light edges. 

Similar to \onlinegreedy, \randomonlinegreedy maintains a set of active servers $S$, along with sets $A(t)$ and $B(t)$. At each time step $t$, the weighted bipartite graph $G_t$ and set of active servers $S$ are passed as input to \greedy, which returns a matching $M_t$. The set $B(t):=\cup_{\tau \leq t}M_\tau$ and $B_i(t)$ represents the set of edges in $B(t)$ connected to server $i$. The set $A_i(t)$ is conditioned on the coin toss for server $i$. If $i\in S_1$, $A_i(t)$ only contains the heavy edges in $B_i(t)$. Otherwise, if $i\in S_2$, $A_i(t)$ only contains the light edges in $B_i(t)$.

At time $t$, if \randomonlinegreedy adds an edge $e=(i,j)$ to $B$, the algorithm checks the weight $W(B_i(t))$ to see if it should be active for the next time step. If $W(B_i(t))>\frac{1}{2}C_i$, then server $i$ is removed from $S$.

The reason for maintaining two sets $B$ and $A$ is that it is possible for $B_i(T)$ to be infeasible for some server $i$. However, $A_i(T)$ is a feasible allocation $\forall\,i$, and $\mathbb{E}\left[W(A_i(T))\right]=\frac{1}{2} W(B_i(T))$. 

The algorithm continues until either $S=\emptyset$ or $t=T$.

\begin{lemma}
The allocation $A_i(T)$ is feasible for each machine $i \in I$.
\end{lemma}


\begin{IEEEproof}
Since \greedy ~ performs a matching at each time step, the degree constraints are always satisfied. We show that the capacity constraints are obeyed as well. Note that $A_i(t) \subseteq B_i(t)$ for all $i$, $t$. By construction, if $W(B_i(t)) > \frac{1}{2}C_i$ at any time $t$, server $i$ is deactivated. Hence every server can accept at most one heavy job. At time $t$, if a server $i\in S_1$ (i.e., it can accept only heavy jobs) is active, there are no heavy edges in $B_i(t-1)$ and the set $A_i(t-1)$ must be empty. If $\exists e=(i,j)\in M(t)$ which is a heavy edge, it is added to $B_i(t-1)$ and $A_i(t-1)$, and server $i$ is deactivated. As $W(B_i(t-1)),\,W(A_i(t-1))$ increase by at most $C_i$ after adding $e$ to $B$ and $A$, it may be that $W(B_i(t))> C_i$ but $W(A_i(t))\leq C_i$ since $A_i(t)$ was empty before. However, if $\exists e=(i,j)\in M_t$ which is a light edge, then it is added to $B_i(t-1)$ but not $A_i(t-1)$, and $A_i(t)$ remains empty. Therefore, if $i\in S_1$, $W(A_i(t))\leq C_i\,\forall\,t$.

On the other hand, if the server $i\in S_2$ is active at time $t$, then $W(A_i(t-1)),W(B_i(t-1))\leq\frac{1}{2}C_i$. If $\exists e=(i,j)\in M_t$ which is a heavy edge, then $e$ is added to $B_i(t-1)$ and $i$ is deactivated. However, $e$ is not added to $A_i(t-1)$ and $W(A_i(t))\leq\frac{1}{2}C_i$ as no edge has been added to $A_i(t-1)$ at time $t$. If $\exists e=(i,j)\in M_t$ which is a light edge, then $e$ is added to $B_i(t-1)$ and $A_i(t-1)$. With the addition of a light edge, $W(B_i(t-1)),W(A_i(t-1))$ increase by at most $\frac{1}{2}C_i$, and as $W(A_i(t-1))\leq\frac{1}{2}C_i,\;W(A_i(t))\leq C_i$. Therefore, if $i\in S_2$, $W(A_i(t))\leq C_i\,\forall\,t$. 
\end{IEEEproof}

\vspace*{0.1in}


\begin{exmp}\label{example: example for infeasibility of B}
This example illustrates how $B_i(T)$ may be an infeasible allocation, while $A_i(T)$ is feasible. Consider a single server with capacity $C$. At each time step, one job is presented, and $T=2$. At $t=1$, a job of weight $\frac{C}{2}-\epsilon$ is presented, while at time $t=2$, a job of weight $C$ is presented. \randomonlinegreedy will put both jobs into $B(2)$. If the coin showed heads, $A(2)$ will contain the second edge. If the coin showed tails, $A(2)$ will contain the first edge at time $t=1$, i.e., $A(2)=\{\frac{1}{2}C-\epsilon\}$ or $A(2)=\{C\}$, and both allocations occur with probability $\frac{1}{2}$. However, $W(B(2))=\left(\frac{3}{2}C-\epsilon\right)$, which is an infeasible allocation.
\end{exmp}

\begin{exmp}\label{example: example for performance of randomonlinegreedy on bad example}
This example illustrates how \randomonlinegreedy performs well on Example \ref{example: deterministic adversarial setting}. In the deterministic setting, accepting/rejecting a job at time $t=1$ can cause an arbitrarily bad competitive ratio because the adversary has freedom to choose the weights in the next time step. The key idea behind the randomization in \randomonlinegreedy is that if there are heavy jobs in the future, then no assignment of light jobs should be made until the heavy jobs are presented, and \randomonlinegreedy does this with probability 0.5. Similarly, if there are no heavy jobs in the future, then light jobs must be assigned to the server and \randomonlinegreedy does this with probability 0.5.

Consider the allocation made by \randomonlinegreedy on the sequence in Example \ref{example: deterministic adversarial setting}. If the job weights to server $i$ are $C$ at $t=2$, then the optimal matching decision would be to not make any allocations to server $i$ at $t=1$, an event which occurs in \randomonlinegreedy with probability 0.5 (i.e., if $i\in S_1$). Similarly, if the job weights to server $i$ are 0 at $t=2$, then the optimal matching decision would be to allocate a job of weight $\epsilon$, an event which occurs in \randomonlinegreedy with probability 0.5 (i.e., if $i\in S_2$). Thus, for the sequence in Example \ref{example: deterministic adversarial setting}, with probability 0.5, \randomonlinegreedy finds the optimum allocation for a server. 
\end{exmp}

\begin{theorem}\label{theorem: random online greedy theorem}
\randomonlinegreedy  is $6-$competitive.
\end{theorem}
\begin{IEEEproof}
Let $W(A^*(T))=W(\cup_{i=1}^n A_i^*(T))$ be the value of the allocation made by the optimal offline algorithm, and $W(B(T))=W(\cup_{i=1}^n B_{i}(T))$ be the value of the infeasible allocation $B(T)$. Moreover, let $\mathbb{E}\left[W(A)\right]=\mathbb{E}\left[W(\cup_{i=1}^n A_{i}(T))\right]$ be the expected value of the feasible allocation $A(T)$ made by \randomonlinegreedy(denoted as $\mathcal{A}$), then from \textbf{Lemma \ref{lemma: random online greedy M1 lemma}} and \textbf{Lemma \ref{lemma: random online greedy M2 lemma}},
\begin{displaymath}
\mu(\mathcal{A})=\min\left(\dfrac{\mathbb{E}\left[W(A(T))\right]}{W(A^*(T))}\right)=\dfrac{1}{6}.\\
\end{displaymath}
\end{IEEEproof}
\vspace*{0.1in}
\begin{lemma}\label{lemma: random online greedy M1 lemma}
\begin{displaymath}
\dfrac{W(B(T))}{W(A^*(T))}\geq \dfrac{1}{3}.
\end{displaymath}
\end{lemma}
\vspace*{0.1in}
\begin{IEEEproof}
As the arguments for (\ref{equation: 3comp sum i not in S}), (\ref{equation: 3comp sum i in S}) hold for the sets $B_i(t)\,\forall\,i$, the proof for \textbf{Lemma \ref{lemma: random online greedy M1 lemma}} follows the same lines as the proof for \textbf{Theorem \ref{theorem: online greedy theorem}}. A full proof is provided in the Appendix.
\end{IEEEproof}
\vspace*{0.1in}
\begin{lemma}\label{lemma: random online greedy M2 lemma}
\begin{displaymath}
\dfrac{\mathbb{E}\left[W(A(T))\right]}{W(B(T))}=\dfrac{1}{2}.
\end{displaymath}
\end{lemma}
\vspace*{0.1in}
\begin{IEEEproof}
The set $B_{i}(t)$ can be partitioned into two mutually exclusive subsets $X_{i}(t)$ and $Y_{i}(t)$, such that $X_i(t)$ only contains heavy edges, while $Y_i(t)$ only contains light edges. Note that $|X_i(t)|\leq 1$. Let $v_i=1 (=0)$ if server $i$ accepts only heavy (light) jobs. As $A_i(t)$ is a feasible allocation $\forall\,t$ and $A_i(t)=X_i(t),t\leq T$ if $v_i=1$, and $A_i(t)=Y_i(t),t\leq T$ if $v_i=0$, $X_i(t),Y_i(t),t\leq T$ are both feasible allocations.

Therefore, 
\begin{eqnarray}
B_i(t)&=&X_i(t)\cup Y_i(t),\;X_i(t)\cap Y_i(t)=\emptyset\,\forall\,t\nonumber\\
W(B_i(t))&=&W(X_i(t))+W(Y_i(t)).\nonumber
\end{eqnarray}
Hence
\begin{eqnarray}
\mathbb{E}\left[W(A_i(T))\right]&=&\mathbb{P}\left[v_i=1\right]W(A_i(T)\mid v_i=1)+\nonumber\\
&&\mathbb{P}\left[v_i=0\right]W(A_i(T)\mid v_i=0),\nonumber\\
&=&\frac{1}{2}\left(W(X_i(T))+W(Y_i(T))\right).\nonumber
\end{eqnarray}
Therefore,
\begin{displaymath}
\mathbb{E}\left[W(A_i(T))\right]=\frac{1}{2}W(B_i(T)).
\end{displaymath}
Summing over all servers $i$,
\begin{eqnarray}
\label{equation: final step of random online greedy M2 lemma}
\sum_{i=1}^{n}\mathbb{E}\left[W(A_i(T))\right]&=&\frac{1}{2}\sum_{i=1}^{n}W(B_i(T)),\nonumber\\
\dfrac{\mathbb{E}\left[W(A(T))\right]}{W(B(T))}&=&\dfrac{1}{2}.
\end{eqnarray}
\end{IEEEproof}

\subsection{Deterministic Algorithm for Small Job Weights and Parallel Servers}


Servers are \emph{parallel} if $C_i = C_{i'}$ and $e_{ij} = e_{i'j}$ for all jobs $j$ and all servers $i$, $i'$. That is, the servers are identical, and each job consumes the same quantity of resources on each server. Thus instead of edge weights we now refer to the weight of each job. If servers are parallel, each with capacity $C$, and each job has weight at most $\epsilon$, then we show a simple deterministic load-balancing algorithm that is $\dfrac{1}{1 - 2\epsilon/C}$-competitive.

\begin{algorithm}[h]
\label{algorithm: parallel}
\caption{\parallelalgo}
\SetKwInOut{Input}{Input}\SetKwInOut{Output}{Output}
\Input{Capacities $C$ of servers \\
	Jobs $J(t)$ at each time step $t \in \{1, \dots, T\}$, \\
	with weight $w(j)$ for $j \in J(t)$.}
\Output{Feasible server allocations $A_{i}$, $i\in\{1,2,...,n\}$}
\Begin{
$A_i \leftarrow \emptyset$ $\forall i \in \{1, \dots, n\}$ initially. \\
\For{$t\leftarrow1$ \KwTo $T$}{
	\For{$j \in J(t)$, \emph{in decreasing order of weight}} {
		\vspace*{0.05in}
		Let $i$ be the machine with highest remaining capacity $C - W(A_i)$ that is not assigned a job in current time step. \\
		\If{$W(A_i \cup \{j\}) \le C$}{
			$A_i \leftarrow A_i \cup \{j\}$ \\
		}
		\Else \Return
	}
}
}
\end{algorithm}

\begin{lemma}
After any time step $t$, the remaining capacity of any pair of machines $i$, $i'$ differs by at most $\epsilon$ with the \parallelalgo algorithm.
\label{lem:capacitydiff}
\end{lemma}

\begin{IEEEproof}
The proof is by induction. Suppose the lemma is true at the end of time step $t-1$, and $A_i(t-1)$, $A_{i'}(t-1)$ are the set of jobs assigned by the algorithm to machines $i$, $i'$ until time step $t-1$. Assume without loss of generality that $W(A_i(t-1)) \le W(A_{i'}(t-1))$. Then by the inductive hypothesis, $W(A_i(t-1)) \ge W(A_{i'}(t-1)) - \epsilon$. Further if $j$, $j'$ are the jobs assigned to $i$, $i'$ respectively in time step $t$, then by the algorithm $\epsilon \ge w(j) \ge w(j')$. It follows that $|W(A_i(t)) - W(A_{i'}(t))| \le \epsilon$.
\end{IEEEproof}

\begin{theorem}
Algorithm \parallelalgo is $\dfrac{1}{(1-2\epsilon/C)}$-competitive.
\end{theorem}

\begin{IEEEproof}
If the \textbf{else} condition in \parallelalgo is never encountered, then at every time step the $n$ jobs of largest weight are assigned, and hence the assignment obtained is optimal. Suppose that for some time step $t$, job $j$, and machine $i$, the \textbf{else} condition is encountered. Thus $W(A_i \cup \{j\}) > C$, and since each job has weight at most $\epsilon$, $W(A_i) \ge C - \epsilon$. By Lemma~\ref{lem:capacitydiff}, for any machine $i'$, $W(A_{i'}) \ge C - 2 \epsilon$. The proof immediately follows.
\end{IEEEproof}

\section{Conclusions}
In this paper, we have derived a $6$-competitive randomized online algorithm for solving the OBRM problem, that generalizes some well studied problems, and is relevant for many applications. There has been a large body of work on special cases of OBRM (adwords problem) or instances of OBRM, where  constant factor competitive algorithms have been derived, however, when weights are small and stochastic or in the randomized input model. Our results in contrast are valid for any arbitrary input, and thus generalize the prior work in a fundamental way. We also expect our ideas to apply for more general instances of online GAP.

\bibliographystyle{IEEEtran}
\bibliography{Reference, myref2}

\section{APPENDIX}
\subsection{Proof of Lemma \ref{lemma: random online greedy M1 lemma}}

\begin{IEEEproof}
For each time step $t$, let $M(t)$ denote the matching produced by \randomonlinegreedy, and let $M^*(t)$ denote the corresponding matching given by the optimal offline algorithm. Let $A^*(t) = \cup_{\tau \leq t} M^*(\tau)$, and $A_i^*(t)$ is the set of edges to server $i$ in the optimal allocation until time $t$. Also, $A_i^*=A_i^*(T)$ and $B_i=B_i(T)$. 

For each edge $(i,j) \in M^{*}(t)\setminus M(t)$, there are three possible reasons why the edge was not selected by \randomonlinegreedy:
\begin{enumerate}
\item An edge $f=(i,j') \in M(t),\,j'\neq j$ \textit{blocks} $(i,j)$, i.e. server $i$ was matched to some job $j'$ by \greedy, such that $w(i,j')\geq w(i,j)$.


\item An edge $f=(i',j) \in M(t),\,i'\neq i$ \textit{blocks} $(i,j)$, i.e. job $j$ was matched to some server $i'$ by \greedy, such that $w(i',j)\geq w(i,j)$.


\item The server $i$ was inactive at time step $t$, i.e., $i\notin S$. 

\end{enumerate}

Let $E_{1}(t)$, $E_{2}(t)$ and $E_{3}(t)$ denote the set of edges in $M^*(t) \setminus M(t)$ that satisfy the first, second and third condition respectively. Clearly, $E_{1}(t)\cup E_{2}(t)\cup E_{3}(t) =M^{*}_{t}\setminus M_{t}$. \textit{Note:} No edge can satisfy the first and third condition simultaneously, as a server which is inactive at time $t$ cannot be matched to any job at time $t$. Therefore, $E_{1}(t)\cap E_{3}(t)=\emptyset$. However, in general, $E_{1}(t)\cap E_{2}(t)\neq\emptyset$ and $E_{2}(t)\cap E_{3}(t)\neq\emptyset$, as edges can satisfy conditions 1 and 2 or 2 and 3.

\vspace*{0.1in}

For all servers $i, i\notin S$, since $W(A^*_i)\leq C_i$ and $W(B_i)> \frac{1}{2}C_i$, the allocation $B_i$ is a $\frac{1}{2}$ approximation to $A_i^*$, i.e.,
\begin{eqnarray}\label{equation: 3comp randomonlinegreedy sum i not in S}
\sum_{i:i\notin S}\sum_{e\in A_i^*}w(e)< 2\sum_{i:i\notin S}\sum_{e\in B_i}w(e)\,.
\end{eqnarray}

\vspace*{0.1in}

Let $E_1=\cup_{t=1}^{T}E_1(t),E_2=\cup_{t=1}^{T}E_2(t),E_3=\cup_{i=1}^{T}E_3(t)$.  Define $E_1^S=\{e=(i,j)\in E_1\mid i\in S\},E_2^S=\{e=(i,j)\in E_2\mid i\in S\}$. Clearly, $E_1^S\cup E_2^S=\cup_{i:i\in S}\left(A_i^*\setminus B_i\right)$, as no edge $e=(i,j),i\in S$ can satisfy the third condition.

The edges $e\in E_1^S\cup E_2^S$ were not selected in the greedy allocation as they were blocked by edges of heavier weight from $B\setminus A^*$. The edges in the set $B\setminus A^*$ are of two types:
\begin{enumerate}
\item $f=(i,j)\in B_i\setminus A_i^*, i\in S$. As all edges $e=(i',j')\in E_1^S\cup E_2^S$ are such that $i'\in S$, $e$ was blocked either because $e$ and $f$ share a server vertex ($i=i'$) or they share a job vertex ($j=j'$). Thus, for every edge $f=(i,j)\in B_i\setminus A_i^*,i\in S$, there may exist at most two edges $e_1=(i,j'),e_2=(i',j)$  such that $e_1,e_2\in E_1^S \cup E_2^S$ and $w(f)\geq w(e_1),w(f)\geq w(e_2)$.

\item $g=(i,j)\in B_i\setminus A_i^*, i\notin S$. As all edges $e=(i',j')\in E_1^S\cup E_2^S$ are such that $i'\in S$, $e$ was blocked only because $g$ and $e$ share the same job vertex ($j=j'$) and $g$ was greedily picked first. Thus, for every edge $g=(i,j)\in B\setminus A^*,i\notin S$, there may exist at most one edge $e_1=(i',j)\in E_1^S\cup E_2^S$ such that $w(g)\geq w(e_1)$.
\end{enumerate}

As $f=(i,j)\in B_i\setminus A_i^*, i\in S$ can block at most two edges in $E_1^S \cup E_2^S$ and $g=(i,j)\in B_i\setminus A_i^*, i\notin S$ can block at most one edge in $E_1^S \cup E_2^S$,
\begin{eqnarray}\label{equation: 3comp randomonlinegreedy sum i in S}
\sum_{i:i\in S}\sum_{e\in A^*_i\setminus B_i}w(e)&=&\sum_{e\in E_1^S \cup E_2^S}w(e)\nonumber\\
&\leq&2\sum_{i:i\in S}\sum_{f\in B_i\setminus A_i^*}w(f)\nonumber\\
&&+\sum_{i:i\notin S}\sum_{g\in B_i\setminus A^*_i}w(g)\,.
\end{eqnarray}

Adding (\ref{equation: 3comp randomonlinegreedy sum i not in S}), (\ref{equation: 3comp randomonlinegreedy sum i in S}),
\begin{displaymath}
\sum_{i:i\notin S}\sum_{e\in A_i^*}w(e) + \sum_{i:i\in S}\sum_{e\in A^*_i\setminus B_i}w(e)\leq 2\sum_{i:i\notin S}\sum_{e\in B_i}w(e) +
\end{displaymath}
\begin{displaymath}
2\sum_{i:i\in S}\sum_{f\in B_i\setminus A_i^*}w(f) + \sum_{i:i\notin S}\sum_{g\in B_i\setminus A^*_i}w(g).
\end{displaymath}

Adding $\sum_{i:i\in S}\sum_{e\in B_i\cap A_i^*}w(e)$ to LHS and RHS,
\begin{displaymath}
\sum_{i:i\notin S}\sum_{e\in A_i^*}w(e) +\sum_{i:i\in S}\sum_{e\in A^*_i}w(e)\leq\sum_{i:i\in S}\sum_{e\in B_i\cap A_i^*}w(e)
\end{displaymath}
\begin{displaymath}
+2\sum_{i:i\in S}\sum_{f\in B_i\setminus A_i^*}w(f)+3\sum_{i:i\notin S}\sum_{g\in B_i}w(g).
\end{displaymath}

Simplifying,
\begin{equation}\label{equation: 3comp randomonlinegreedy final equation}
\sum_{i\in I}\sum_{e\in A_i^*}w(e)\leq 3\sum_{i\in I}\sum_{e\in B_i}w(e).\\
\end{equation}
\vspace*{0.1in}
\end{IEEEproof}

\end{document}

%% file: introduction.tex
\section{Introduction}

We consider a basic combinatorial online resource allocation problem, where there are $n$ servers with capacity $C_i$, $i=1,\dots, n$. At time step $t$, a set of jobs arrive, where job $j$ has weight $w(i,j)$ on server $i$. Thus, the graph $G(t)$ consisting of a set of job vertices and weighted edges incident to the server vertices  
is revealed at time $t$.
The problem is to assign these jobs to the servers so that each server is assigned at most one job and each job is assigned to at most one server in each time step. Further, the set of jobs assigned to each server must respect the capacity constraints for the server. That is, the sum of weight of jobs assigned to server $i$ over all time steps 
has to be at most $C_i$. Thus, essentially, a matching has to be found at each time step, that also respects the capacity constraints. In general, a job may have different weights depending on the server it is assigned to. The allocation has to be done irrevocably at each time $t$ in an online fashion, i.e., without the knowledge of jobs and edges that arrive in the future. 
Let the sum of weights assigned to server $i$ at the end of input sequence be $W_i$. 
Then the objective is to maximize the sum $\sum_{i=1}^{n} W_i$. We call this the online budgeted repeated matching (OBRM) problem. 

To characterize the performance of an online algorithm, we consider the metric of competitive ratio that is defined as the ratio of the profit made by the online policy and the offline optimal policy, minimized over all input sequences. The competitive ratio is a worst case guarantee on the performance of an online algorithm. An online algorithm is called $\gamma$-competitive if its worst case competitive ratio is at least $1/\gamma$. 

Suppose the graphs $G(t), t \ge 0$ are revealed ahead of time, i.e., if we consider the offline setting, then the above described budgeted repeated matching (BRM) is an instance of a generalized assignment problem (GAP) \cite{fleischer2006tight, shmoys1993approximation}. A GAP is defined by a set
of $n$ bins and a set of $m$ items to pack in the bins, with `value' $v_{ij}$ 
for assigning item $j$ to bin $i$. In addition, there are constraints for each bin (possibly size constraints), describing which subset of items can fit in that bin. The objective is to maximize the total value of items packed in the bins, subject to the bin constraints. Our problem introduces an additional constraint that the bin assignment must obey matching constraints. As the LP in \cite{fleischer2006tight} allows for a feasible set which is exponential in the size of the input, the additional constraint does not violate any conditions and our problem is a valid instance of the GAP. For the GAP, \cite{fleischer2006tight} derived a $\beta \left(1-\frac{1}{e}\right)$-approximate offline algorithm, where $\beta$ is the best approximation ratio for solving the GAP with a single bin, improving upon special case results of \cite{shmoys1993approximation}. The online GAP has been considered in \cite{alaei2013online}, however, with two restrictions; that the weights and sizes of each item are stochastic and each items' size is less than a fixed fraction of the bin capacity.

There are many motivations for considering the OBRM, we list the most relevant ones below.

\noindent {\bf $\mathsf{adwords}$ problem}: The classical $\mathsf{adwords}$ problem \cite{mehta2007adwords} is a special case of OBRM, where at each time slot, only one job (called impression) arrives, and each of the $n$ advertisers reveal their preferences or bids for each impression, that defines how much that advertiser is willing to pay to the platform for displaying his/her ad for that impression.  
Each advertiser has a total budget of $B_i, i=1,\dots, n$ and the problem is to assign one advertisement to each impression, so as to maximize the overall revenue (sum of the weights on the assigned edges across the advertisers), subject to each advertiser's budget constraint. 

A more general problem is the $\mathsf{adwords}$ problem with multiple slots \cite{buchbinder2007online}, where there are multiple slots for which an advertiser can bid for each impression,  and the advertiser preferences depends on the slot index. 
In particular, for each impression $j$, a graph is revealed between the advertisers and slots with bids/weights $b(i,s)$ for advertisers $i=1,\dots, n$ and slots $s=1,\dots,L$, and the problem is to find a matching for each impression across multiple slots, so as to maximize the overall revenue subject to advertisers' individual budget constraints. 
It is easy to see that $\mathsf{adwords}$ problem with multiple slots is an OBRM, 
where time slots have been replaced by each impression, and there at most $L$ jobs at each time. 

\noindent {\bf Caching problem}: To serve the ever increasing video traffic demand over the internet, many Video on Demand (VoD) services like Netflix \cite{Netflix} and Youtube \cite{Youtube} use a two-layered content delivery network \cite{Netflix_openconnect}. The network consists of a back-end server which stores the entire catalog of contents offered by the service and multiple front-end servers, each with limited service and storage capacity, located at the `edge' of the network, i.e., close to the end users \cite{SGSS14, LLM12,XT13,LLM13,aaglr10}. Let $m$ be the total number of different contents/files that can be accessed by any user. Let $n$ be the front end servers or caches as they are popularly called, with capacities $C_i$, and let server $i$ store subset $S_i \subseteq [m]$. Thus, at any time, each server can at most serve $C_i$ requests, for any of the contents belonging to $S_i$. At each time slot, multiple content access requests arrive, and the problem is to map (match) these requests to different servers so as to maximize the total number of served requests \cite{HongHou, SGSS14, LLM12,XT13,LLM13,aaglr10}. A request not served is assumed to be dropped. Thus, equivalently, we want to minimize the number of dropped requests subject to individual server capacity constraints. It is easy to see that this caching problem is an instance of the OBRM.

\noindent {\bf Scheduling problem}: OBRM can be considered as a scheduling problem on parallel machines, where each machine has a total capacity, and each job has a profit associated with each machine. At each time, a set of jobs arrive to the scheduler, and the problem is to find an online matching of jobs, that maximizes the total profit subject to machine capacity constraints. 

\noindent {\bf Crowdsourcing}: A problem of more recent interest is the crowdsourcing problem \cite{ho2012online}, where there are $n$ tasks that need to be accomplished and 
each has individual budgets $b_i$. User/worker $j$ on its arrival, reveals a utility $u_{ij}$, and the goal is to match workers with jobs that maximize the utility, subject to the budget constraints. An extension to this problem where more than one worker arrives at the same time, and each worker can only be assigned at most one job is equivalent to OBRM.

\subsection{Contributions}
We make the following contributions in this paper.
\begin{itemize}
\item We propose a simple greedy algorithm for OBRM that is shown to be $3$-competitive, whenever the weight of any edge is at most half of the corresponding server capacity. As will be shown later (Example \ref{example: deterministic adversarial setting}), no deterministic algorithm has bounded competitive ratio in the unrestricted case when the edge weights are arbitrary. Thus, some restriction on the edge weights is necessary. In fact, we prove a more general result that if the weight of any edge is at most $\alpha$ times the corresponding server capacity, the greedy algorithm is $\left(1+\frac{1}{1-\alpha}\right)$-competitive. We show via an example that our analysis of the algorithm is tight.
\item For the unrestricted edge weights case, we propose a randomized version of the greedy algorithm and show that it is $6$-competitive when the edge weights are arbitrary against an \emph{oblivious} adversary, that decides the input prior to execution of the algorithm. That is, the adversary decides the input before the random bits are generated. For our algorithm, we define a job as \emph{heavy} for a server if its weight is more than half of the server capacity, and \emph{light} otherwise. Our randomization is rather novel, where a server accepts/rejects heavy jobs depending on a coin flip. Typically, the randomization is on the edge side, where an edge is accepted or not depending on the coin flips.
\item Finally, when each server has identical capacity $C$, and is \emph{parallel}, that is, a job has the same weight on every server, we give a deterministic $C/\left(C-2\epsilon\right)$-competitive algorithm, where $\epsilon$ is the maximum job weight. Thus if $\epsilon/C \rightarrow 0$, this algorithm is nearly \emph{optimal}.

\end{itemize}
\subsection{Related Work and Comparison}
There is a large body of work on problems closely related to OBRM, specifically, the $\mathsf{adwords}$ problem or the budgeted allocation problem \cite{mehta2007adwords}, $\mathsf{adwords}$ problem with multiple slots \cite{buchbinder2007online}, the offline budgeted allocation problem \cite{azar2008improved, srinivasan2008budgeted}, the stochastic budgeted allocation problem \cite{devanur2009adwords, feldman2010online, haeupler2011online, mehta2012online}, secretarial knapsack problem \cite{babaioff2007knapsack}. We describe them in detail, and contrast the prior results with ours to put our work in the right perspective.

In the offline scenario, a constant factor $3/2$ approximation ratio \cite{azar2008improved} is known for the budgeted allocation problem, when only one job arrives at each time, which was later improved to $4/3$ in \cite{srinivasan2008budgeted} that meets the lower bound from \cite{andelman2004auctions}. For the offline scenario, approximation ratio is defined as the ratio of the profit of a poly-time algorithm and the optimal algorithm.

A common theme in solving the 'online' budgeted matching/allocation problem with arbitrary inputs is to assume that the weight of any edge is `small' compared to the respective server capacity. Under this assumption, starting with \cite{mehta2007adwords}, almost close to optimal online algorithms have been derived in \cite{mehta2007adwords, buchbinder2007online} using many different ideas such as $\psi$ functions, PAC learning and primal-dual algorithms. However, for all these algorithms, a constant factor competitive ratio is only 
possible if the weight of any edge is vanishingly small compared to respective server capacity, which otherwise grows with the largest ratio of any edge weight and the respective server capacity. 

For stochastic input with known distribution, OBRM with single job arrival at each time has also been studied extensively in literature \cite{devanur2009adwords, feldman2010online, haeupler2011online, mehta2012online}. Assuming small edge weight, \cite{devanur2009adwords, feldman2010online} achieve near optimal $1+o(\epsilon)$ competitive ratio, while \cite{mehta2012online} gives a $1/.567$ competitive ratio. The case when estimates are unreliable has been studied in \cite{mahdian2007allocating}.

From a resource allocation or crowdsourcing job matching perspective, OBRM with single job and stochastic input has been studied in \cite{tan2012online, jaillet2012near} and \cite{ho2012online}. In a minor departure from other work, \cite{tan2012online} allowed a little bit of slack in terms of capacity constraint and showed that the derived profit is within a $O(\epsilon)$ of the optimal profit while allowing constraint violations of $O(1/\epsilon)$. For caching applications, assumptions are made either on the the small-job sizes  \cite{HongHou} or on large number of servers and asymptotic results \cite{HongHou, SGSS14, LLM12,XT13,LLM13,aaglr10} are found.

The most general relevant result is the $\frac{1}{1-\frac{1}{\sqrt{k}}}$-competitive algorithm in expectation for the online GAP \cite{alaei2013online}, where similar to prior work two restrictions are made; that the weights and sizes of each item are stochastic and each items' size is less than a $\frac{1}{k}$ of the bin capacity. 

To put our results in perspective, for the deterministic algorithm, we do not make small jobs assumptions, and allow each edge weight to be as large as half the corresponding server capacity. Thus, for our scenario, when we assume that only one job arrives each time, the algorithms of \cite{mehta2007adwords, buchbinder2007online} will give a non-constant competitive ratio. Similar result will be obtained from \cite{buchbinder2007online} when multiple jobs arrive at the same time, using the $\sf{adwords}$ with multiple slots problem solution. 
To overcome the edge weight restriction with respect to server capacities, rather than randomizing the input as done in works with stochastic weights, we randomize the algorithm, where each server tosses an unbiased coin once independently, before the start of the input. A server whose toss comes heads, only accepts a job (edge selected by greedy algorithm) if it is at least as much as half of its capacity, while otherwise, the server accepts a job only if it is at most half of its capacity. The basic idea behind this randomization is that with probability half, the sum weight of all accepted jobs by any server is at least as much as the sum weight obtained at that server by the greedy algorithm under the restricted weights setting. Thus, completely eliminating the need to restrict the job size (edge weights), and get a $6$-competitive randomized online algorithm with worst case input.

To highlight the fact that we do not have to consider the randomized input setting, we discuss two related problems where the input has to be randomized to get non-trivial competitive ratios. First is the online matching problem with no capacity constraint, where at each time one job arrives and has to be assigned irrevocably to one of the servers, and once a server is allotted one job, it cannot be matched to any subsequently arriving job. The best known algorithm for online matching is $8$-competitive \cite{korula2009algorithms}, under a randomized input model, that improves upon earlier works of \cite{dimitrov2012competitive, babaioff2007matroids}.
For the online matching problem, since there is no constraint on edge weights of each job, the competitive ratio of any algorithm with the arbitrary input is unbounded. Hence to overcome this degeneracy, a randomized input model is considered, where the weights can possibly be chosen by an adversary but the order of arrival of jobs is uniformly random.
In contrast, with OBRM problem, since each edge weight is at most equal to the capacity of the corresponding server, we can get a $6$-competitive randomized algorithm even in the worst case input model.

The second problem is a special case of online GAP is an online knapsack problem \cite{babaioff2007knapsack}, where there is a single bin 
with capacity $C$. In each time slot, a job arrives with value $v$ and space $s$. The problem is to either accept or reject each job irrevocably, so as to maximize the total aggregate value of all accepted jobs subject to the total space of accepted jobs being less than $C$. Even for the online knapsack problem, one has to consider the randomized input model. It is worthwhile contrasting the special case of OBRM when only one job arrives at each time with online knapsack problem. We do not need to randomize the arrival sequence in the OBRM, since both the weight and the space are identical for each job, unlike the online knapsack problem.

In summary, we propose a simple $6$-competitive randomized algorithm for solving an online subclass of GAP with unrestricted weights, that we call OBRM, that has many applications. The problem has been very well studied for the case of single job arrivals, but has escaped general results, and algorithms with constant competitive ratio have been possible only under small-job or stochastic input assumption.